\documentclass[12pt]{article}

\usepackage{scicite}
\usepackage{graphicx}
\usepackage{times}
\usepackage{bm}
\usepackage{color}
\usepackage{amsmath}%
\usepackage{MnSymbol}%
\usepackage{wasysym}%

\newenvironment{sciabstract}{%
\begin{quote} \bf}
{\end{quote}}

\title{Ice nucleation in the wake of warm hydro-meteors}

\author
{Prasanth Prabhakaran,$^{1, 2, 3\ast}$ Gregory Kinney,$^{2}$ Will Cantrell,$^{2}$ \\Raymond A. Shaw,$^{2}$ Eberhard Bodenschatz$^{1, 3, 4}$\\
\\
\normalsize{$^{1}$Max Planck Institute for Dynamics and Self-Organization, 37077 G\"ottingen, Germany.}\\
\normalsize{$^{2}$Michigan Technological University, Department of Physics, Houghton, MI 49931, USA.}\\
\normalsize{$^{3}$Institute for Dynamics of Complex Systems, University of G\"ottingen, 37077 G\"ottingen,}\\
\normalsize{Germany.}\\
\normalsize{$^{3}$Laboratory of Solid State Physics,
Cornell University, Ithaca, NY 14853, USA.}\\
\\
\normalsize{$^\ast$To whom correspondence should be addressed; E-mail:  prasantp@mtu.edu.}
}

\date{}

\begin{document} 


\baselineskip24pt


\maketitle


\begin{sciabstract}
The formation of ice in mixed-phase clouds greatly impacts Earth's hydrologic cycle. %-- most of the precipitation over the continents involves the ice phase. %\cite{mulmenstadt2015, field2015importance}. 
The intensity, distribution and frequency of precipitation as well as radiative properties of clouds in the mid-latitudes are strongly influenced by the number concentration of ice particles. %\cite{vergara2018strong, cantrell2005production}. 
A long-standing riddle in mixed-phase clouds is the frequent observation of measured ice particle concentrations several orders of magnitude higher than measured ice-nucleating particle concentrations. %\cite{review, hobbs1985ice, mossop1985secondary, pruppacher2010cloud, cantrell2005production}. 
%This implies that yet unknown ice nucleation or production mechanism must exist.\cite{review}. 
Here, we report laboratory observations of copious cloud droplets and ice crystals formed in the wake of a warm, falling water drop. Aerosols were activated in the transient regions of very high supersaturation due to evaporative mixing in the wake. %The drop is a laboratory surrogate for a relatively warm hydrometeor (graupel or hail) growing by accretion of super-cooled cloud droplets. 
We extend these results to typical mixed-phase atmospheric conditions, and our calculations show that the induced evaporative supersaturation may significantly enhance the activated ice nuclei concentration in the particle's wake.
\end{sciabstract}

% In setting up this template for *Science* papers, we've used both
% the \section* command and the \paragraph* command for topical
% divisions.  Which you use will of course depend on the type of paper
% you're writing.  Review Articles tend to have displayed headings, for
% which \section* is more appropriate; Research Articles, when they have
% formal topical divisions at all, tend to signal them with bold text
% that runs into the paragraph, for which \paragraph* is the right
% choice.  Either way, use the asterisk (*) modifier, as shown, to

The formation of ice-phase in clouds is responsible for most of the precipitation over the continents \cite{mulmenstadt2015, field2015importance}.
Rain from mixed-phase clouds is very efficient because only a small fraction of supercooled cloud droplets freeze, and the supersaturation with respect to the ice-phase is higher compared to liquid water, resulting in a faster diffusive growth of ice crystals compared to the liquid droplets. The more massive crystals fall relative to the surrounding liquid droplets, resulting in further growth by vapor deposition and collisions with cloud droplets or other ice crystals, thus, rapidly producing precipitation. The efficacy of this process relies heavily on the number of ice particles (IP) \cite{cantrell2005production}. Thus, the generation of additional IP requires activation of more ice nucleating particles (INP) or a secondary ice-production (SIP) mechanism. The process of riming-splintering is the most studied SIP mechanism, however, it is active only in a narrow temperature range and droplet size distribution \cite{hallett1974production, review, mossop1985secondary, cantrell2005production}. Observations show that IP concentrations exceed measured INP concentrations, and quite often occur outside of the parameter space spanned by riming-splintering\cite{review,  hobbs1985ice, mossop1985secondary, pruppacher2010cloud, cantrell2005production}. The role of evaporative supersaturation in the wake of a growing graupel was suggested as a possible ice multiplication mechanism based on an advection-diffusion model of the temperature and vapor field \cite{fukuta1986numerical}. The mixing of the warm water-vapor from the surface of the hydrometeor and the cold cloudy air significantly enhances the supersaturation behind the hydrometeor. This has been recently explored in direct numerical simulations of flow past a growing hail-stone \cite{pc}. A recent experiment in a Sulfur Hexafluoride (SF$_6$)-Helium moist convection system showed that SF$_6$ droplets nucleated homogeneously in the wake of falling cold drops \cite{prabhakaran2017can}. These results proposed that the concept could be of atmospheric relevance, and we explore this question in the present work and show experimentally that it provides, when applied to a hydometeor falling in a cold atmosphere, an important additional ice nucleation mechanism. 

\begin{figure}[]
\centering
\includegraphics[width = 0.95\textwidth]{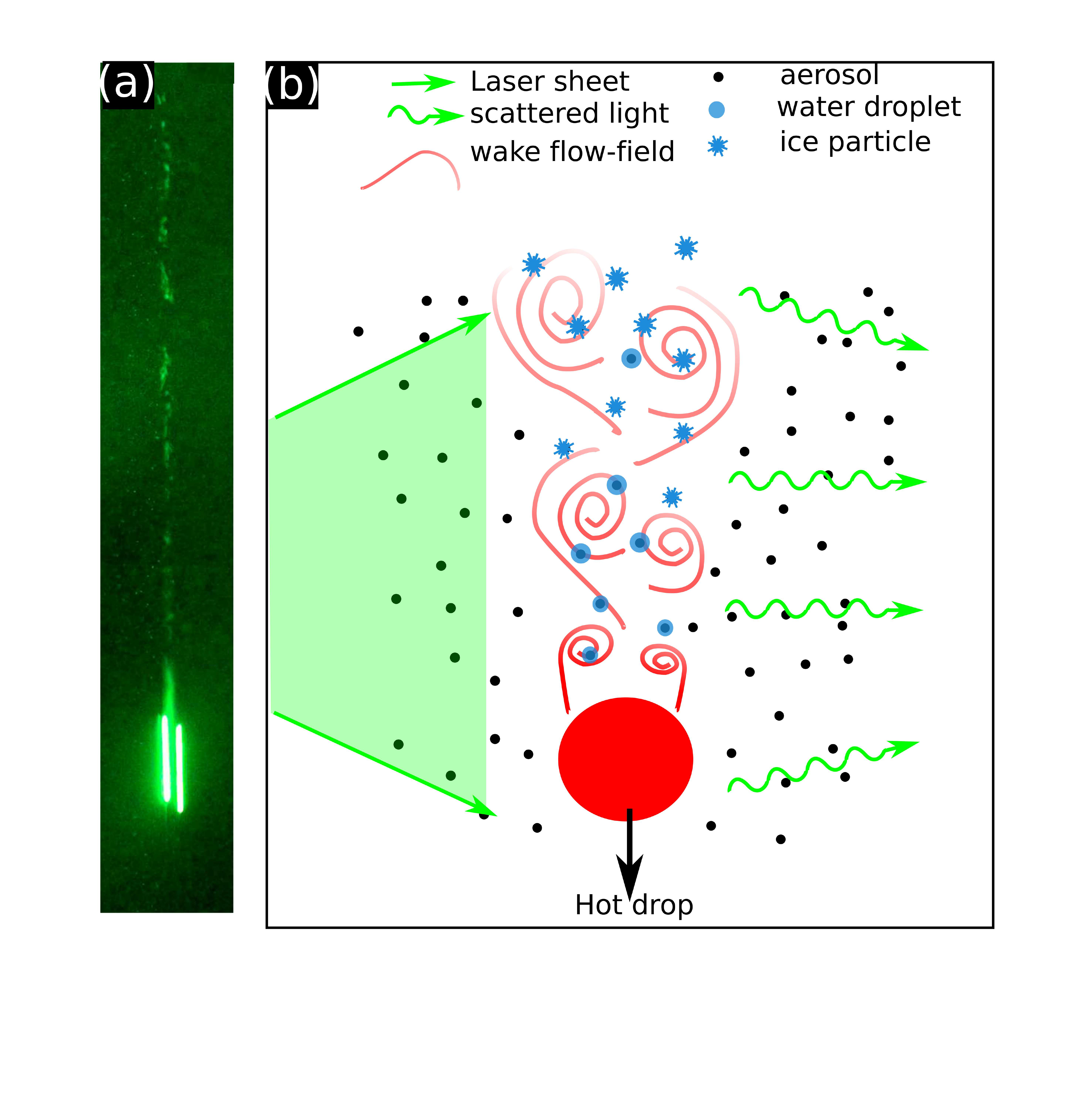}
\caption{Schematic of the experimental setup and the formation of cloud droplets and ice crystals in the wake of a warm falling drop. The warm drop is a surrogate for a hydrometeor that is warmed by the latent heat of fusion as it collects super-cooled cloud droplets in a mixed-phase cloud. (a) A snapshot from the experiment, showing the falling drop (two bright streaks at the bottom due to refraction of the laser light at the back and front water air interfaces of the drop) and a trail of activated ice nuclei and cloud droplets in its wake. See Fig.~\ref{fig:cold_cloud} (a) for details. (b) Schematic of the flow field behind the hot drop. The chamber was seeded with Snowmax aerosols. The intensity of the red color represents the temperature field. As mixing occurs behind the drop, inducing regions of transient, high supersaturation, aerosol particles activate, becoming cloud droplets. A fraction of the nucleated droplets froze in the wake as the injected aerosols were ice-nucleating particles made visible by the glittering reflections from the particles. %The wake of the drop was illuminated using a vertical, planar light sheet from an MGL-III $532$\,nm $300$\,mW continuous laser. The forward scattered light from the nucleated particles was recorded using a Sony alpha 7S II camera placed across the chamber. See supplementary text for details.}
}\label{fig:schematic_ice}
\end{figure}

To that end we ask the question, ``Can we generate a realistic atmospheric environment containing aerosol particles, and visualize their activation and freezing in the wake of a falling hydrometeor that is warmer than ambient?'' In previous experimental studies of flow past growing ice particles such as hailstones, the particle was typically suspended in a column of counter flowing air, which was seeded with supercooled droplets\cite{lesins1986sponginess}. These experiments give insights into particle growth by collision, and allows observation of the near-field of the wake, but are not suitable for investigating the far-field dynamics because the axisymmetric wake rapidly relaxes to ambient conditions\cite{tennekes1972first}. In this work, we change the reference frame to instead observe the evolution of the wake, as a warm water drop falls through a quiescent-cold ambient. This enables us to visualize the far-field dynamics in the wake of the drop. We show here that water droplets and ice particles form in the wake of a falling hot water drop, which is a laboratory surrogate for a relatively warm hydrometeor (graupel or hail) in a cloud growing by accretion of super-cooled cloud droplets. Our experiment is illustrated in  Fig.~\ref{fig:schematic_ice} that shows both the  schematic of the mechanism, as well as, one typical example of ice nucleation in the wake of a falling warm water drop.

\begin{figure}[]
\centering
\includegraphics[height = 0.9\textheight]{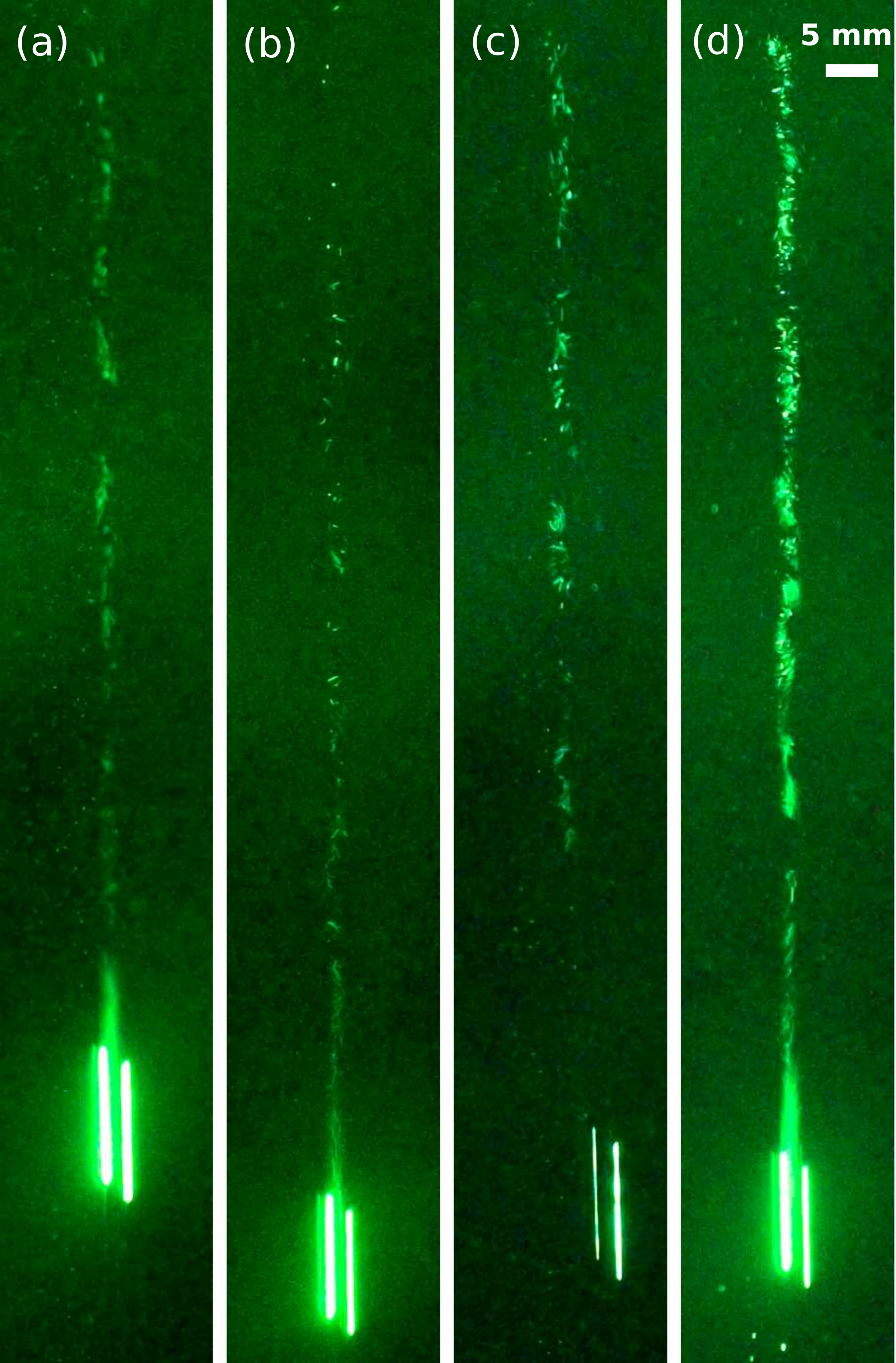}
\caption{Nucleation of water droplets and ice particles in the wake of a hot drop under cold conditions. Ambient temperature $\approx\,-18~^{\circ}$C, relative humidity (RH) $\approx\,60\%$. The initial drop temperatures are (a) $4\,^{\circ}$C (b) $10\,^{\circ}$C (c) $10\,^{\circ}$C (d) $20\,^{\circ}$C. %(e) $50~^{\circ}$C. 
The number concentration of the aerosol particles in the chamber was about $10^4$ per cm$^{3}$ in all cases except in (b), where it was $\approx~10^3$ per cm$^{3}$.}
\label{fig:cold_cloud}
\end{figure}

Figure%~\ref{fig:warm_cloud} and
~\ref{fig:cold_cloud} illustrates the nucleation dynamics that we see in the wake of a hot drop in cold conditions. (See supplementary material for warm condition experiments.) The wake behind the drop is visible due to the light scattered by the nucleated droplets and IP from the laser sheet. Clearly, particles form, proof that the transient supersaturation is high enough to activate them. Some of the droplets subsequently freeze, which we infer from the pronounced glittering of the scattered light, due to its non-spherical shape, in contrast to the water droplets. The glittering effect was produced only when the temperature was low enough that  some of the nucleated supercooled droplets froze. Note that the spatial extent and the number density of the nucleated particles in the wake increase with the temperature difference between the drop and the ambient (left to right in Fig.~\ref{fig:cold_cloud}). Also note that the nucleated particles do not grow indefinitely as the ambient away from the wake was sub-saturated. If the ambient air were supersaturated, the resulting cloud would obscure the visualization of the dynamics. Furthermore, we note that the concentration of the nucleated particles in the wake is not uniform. There exist pockets of very high concentrations of the particles separated by regions of very low concentrations. This is likely a manifestation of the spatially intermittent nature of entrainment and mixing processes in the wake, which is a hallmark of the free-shear flow \cite{bisset2002turbulent}.

\begin{figure}[]
\centering
\includegraphics[width = \textwidth]{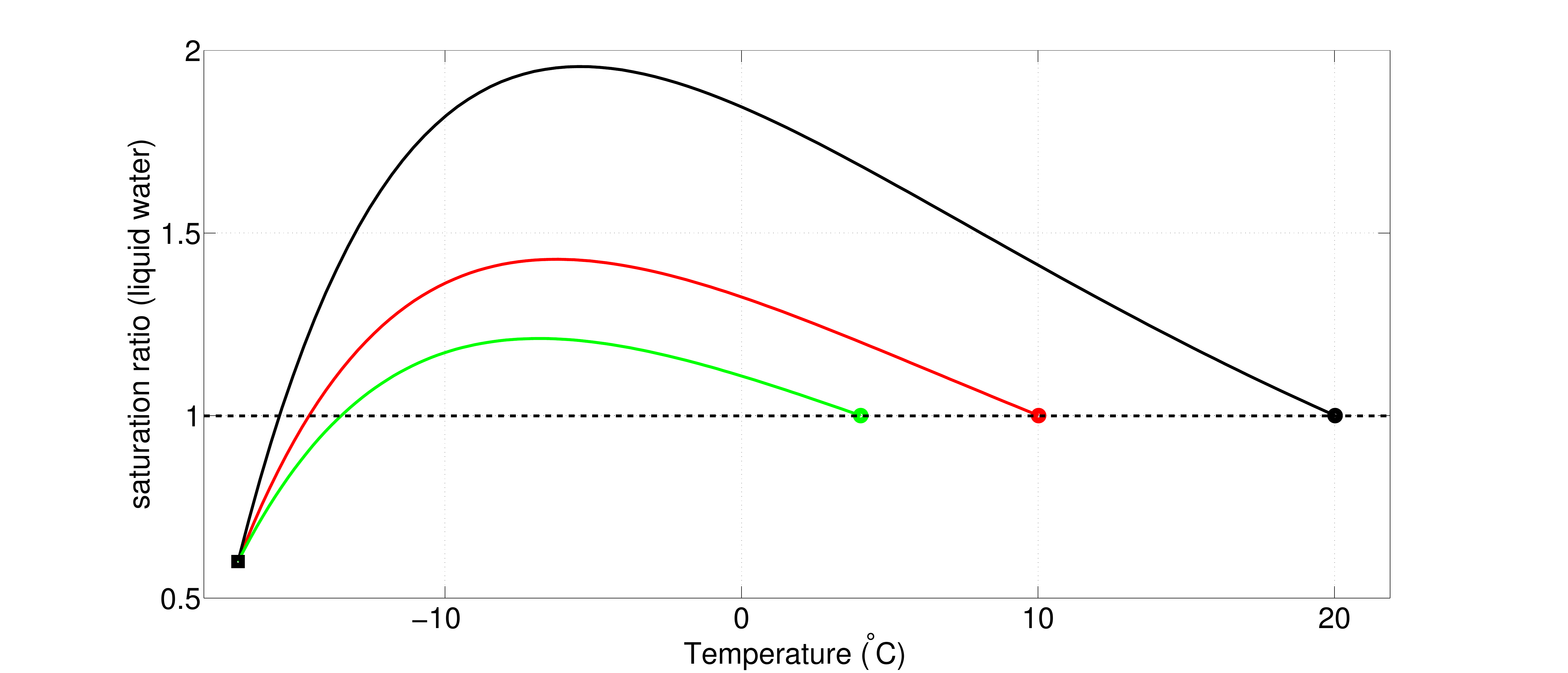}
\caption{Liquid-water saturation ratio $S=\frac{P_w}{P_s}$ for the conditions shown in Fig.~\ref{fig:cold_cloud}. $P_w$ is obtained using Eq.~\ref{eq:mixing_modelP} and $P_s$ is the saturated vapor pressure at $T_w$ (Eq.~\ref{eq:mixing_modelT}). The condition at the surface of the drop is marked using filled circles ($\bullet$) of various colors: $\bullet$ $20~^\circ$C, $\textcolor{red}{\bullet}$ $10~^\circ$C, $\textcolor{green}{\bullet}$ $4~^\circ$C. % $\textcolor{yellow}{\bullet}$ $1~^\circ$C.
The ambient conditions are denoted by $\blacksquare$. The corresponding mixing line, for different mixing fractions $x$ of ambient and wake-influenced air, is represented by a curve connecting $\bullet$ (of different colors) and $\blacksquare$. The horizontal dashed line indicates $S$ = 1.}
\label{fig:cold_ss}
\end{figure}

We use an adiabatic mixing parcel model to estimate the supersaturation in the wake of the warm drop \cite{Bohren_1998}. The temperature and vapor pressure in the wake are then 
\begin{eqnarray}
\label{eq:mixing_modelT}
T_w = xT_a + (1-x)T_d\\
\label{eq:mixing_modelP}
P_w = xP_a + (1-x)P_d
\end{eqnarray}
where $x$ is the mixing fraction between the ambient air and the evaporated vapor from the drop. $T_w$ and $P_w$ are the temperature and vapor pressure of water vapor in the wake. %as a function of mixing fraction $x$.
$T_d$ and $P_d$ are the drop temperature and equilibrium vapor pressure, $T_a$ and $P_a$ are the ambient temperature and vapor pressure. Figure\,\ref{fig:cold_ss} shows how the supersaturation in a fluid parcel initially close to the warm drop varies as a function of the degree of mixing with the ambient air. The mixing fraction $x$ increases from right to left. The supersaturation in the fluid parcel near the drop initially increases with $x$ and above a critical value, it decreases until it attains the ambient condition. We also observe that the evaporative supersaturation attained in the wake increases with the temperature difference between the drop and the ambient. This explains the corresponding increase in the lifetime and the number concentration of the nucleated particles observed in the wake. The experiments discussed here confirm that the transient supersaturation attained in the wake of a warm hydro-meteor can activate water droplets, which subsequently freeze, and that a mixing model captures the essential mechanism reasonably well.

\begin{figure}[]
\centering
\includegraphics[width = \textwidth]{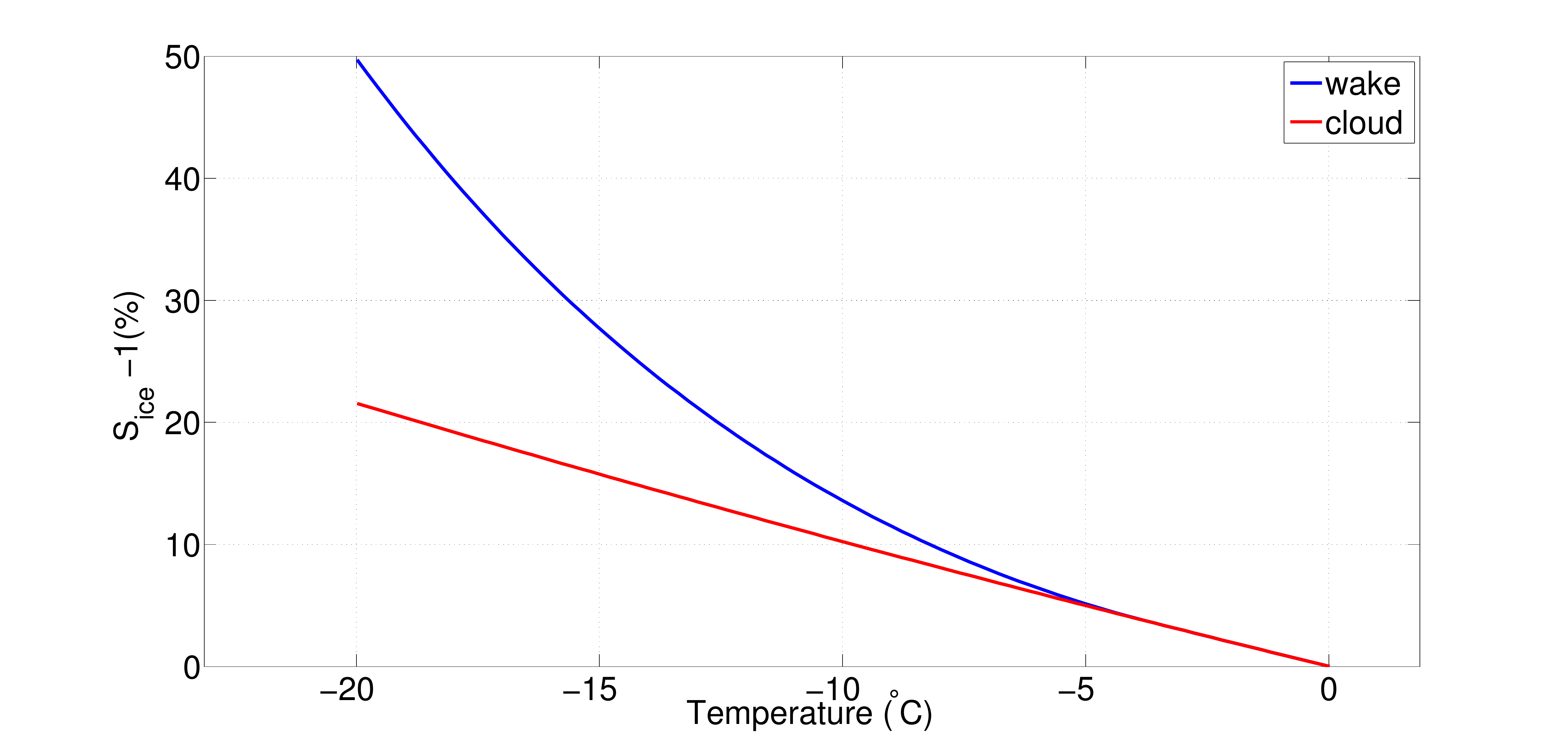}
\caption{Variation of ice supersaturation as a function of ambient temperature under mixed-phase conditions. The red curve represents the ambient ice supersaturation that exists at liquid-water saturation. The blue curve represents the maximum ice supersaturation attained in the wake of a hail/graupel particle in wet-growth mode, obtained from the mixing model.}
\label{fig:ice_enhancement}
\end{figure}

Having established that ice can be formed in the wake of a warm hydrometeor, we turn to the implications for processes in Earth's atmosphere. %As noted above, ice in excess of what can be explained by measured concentrations of ice nucleating aerosol particles is frequently observed \textcolor{blue}{(Maybe this sentence is redundant/out of place)}.
%Ice nucleation processes have been extensively investigated under laboratory conditions and in the field, and are strongly influenced by the environmental temperature and supersaturation with respect to ice \cite{hoose2012heterogeneous}.
The mechanism described above may be relevant to an atmospheric context in that it may make it possible for relatively small particles that would not otherwise become cloud droplets, to do so, and subsequently freeze. Laboratory investigations show that at relatively warm temperatures, the fraction of ice nuclei activated in the condensation/immersion mode, increases with relative humidity in excess of water saturation \cite{hoose2012heterogeneous, koehler2010laboratory}. This is consistent with the fact that smaller aerosol particles require higher supersaturation with respect to water to activate as droplets before they freeze. Moreover, the increase in relative humidity may also increase the probability of deposition nuclei acting as contact nuclei \cite{cooper1974possible}. In the atmosphere, a riming hydrometeor has a surface temperature higher than ambient due to the latent heat released from the freezing process \cite{pruppacher2010cloud}. The temperature difference between the ambient and the hydrometeor surface is directly proportional to the amount of liquid water in the cloud, and in the most extreme scenario, the surface temperature of the riming hydro-meteor is  $0\,^\circ$C \cite{pruppacher2010cloud, yau1996short, rasmussen1987melting, jin2019new}. In this regime, the hydrometeor is said to be in the wet-growth mode because the surface is a layer of liquid water. The warm drop used in the experiments shown in Fig.~\ref{fig:cold_cloud} was a surrogate for this. It should be noted that in the results shown in Fig.~\ref{fig:cold_cloud} the drop temperatures are greater than $0\,^\circ$C. This was required to compensate for the low ambient humidity (RH$\approx60\%$) under laboratory conditions. To extend our laboratory results to the cloud context, we consider a hailstone in the wet growth regime falling through a convective cloud with a supersaturation of $0.1\%$ \textit{w.r.t} liquid  water. Fig.~\ref{fig:ice_enhancement} shows the ice supersaturation in the cloud as a function of the temperature and the corresponding peak ice supersaturation in the wake of the hailstone obtained from the mixing parcel model. We note that the supersaturation attained in the wake increases with the decrease in the cloud temperature. This enhanced supersaturation may activate very small aerosol particles, which were not active at cloud supersaturation. A fraction of these newly formed cloud droplets may also freeze in the immersion mode.

As an initial estimate, we assume that the cloud condensation nuclei ($CCN$) number concentration as a function of supersaturation is represented as $N_{CCN}=Cs^k$, where $s$ is the supersaturation in percent and $C$ and $k$ are parameters that vary depending on the geographical location \cite{twomey1959nuclei, pruppacher2010cloud}.  
For simplicity, we take $k$ to be 1. With that, we estimate a $CCN$ enhancement factor $EF$ defined as the ratio between $N_{CCN}(\textrm{wake})$ and $N_{CCN}(\textrm{cloud})$. Let us assume that the supersaturation in the wake is $1\,\%$ and that the supersaturation in the cloud is about $0.1\,\%$. This results in an $EF$ of about 10. If we consider a linear relationship between $N_{CCN}$ and activated INP concentration, then the ice concentration is also increased by a factor of 10 in the wake of a single wet-growing hailstone. Note that $EF$ will increase with the decrease in cloud temperature (see Fig.~\ref{fig:ice_enhancement}). The efficiency for an entire cloud will depend on the nonlinear feedback introduced when ice crystals produced in the wake become graupel particles that produce their own warm, wet wakes.

The discrepancy between the measured concentration of INP and ice in clouds is a long-standing conundrum in cloud physics. It is unlikely to have a single answer; multiple mechanisms will contribute to SIP in different scenarios. Here, we have outlined an ice production process that may self seed deep convective clouds, commonly observed in the tropics (\textit{e.g.}, in the monsoon), leading to the rapid onset of intense precipitation. Our observations and calculations suggest that falling, riming (\textit{i.e.} warm) hydrometers may activate and nucleate small CCN/INP, that otherwise may not become cloud droplets or ice crystals. Additional studies are required to assess the importance of this mechanism for weather and climate predictions. 

%\bibliography{scifile}

\bibliographystyle{Science}

 \section*{Acknowledgments}
We thank K. K. Chandrakar for his help with the camera. \\
\textit{Author contributions:} PP and EB proposed the idea. PP, GK, WC and RAS designed the experiment. PP and GK conducted the experiment. PP, GK, WC, RAS and EB analysed the data. PP, WC, RAS and EB wrote the paper. \\
The authors declare no conflict of interest.\\

%Here you should list the contents of your Supplementary Materials -- below is an example. 
%You should include a list of Supplementary figures, Tables, and any references that appear only in the SM. 
%Note that the reference numbering continues from the main text to the SM.
% In the example below, Refs. 4-10 were cited only in the SM.     
\section*{Supplementary materials}
\begin{figure}
\centering
\includegraphics[width = \textwidth]{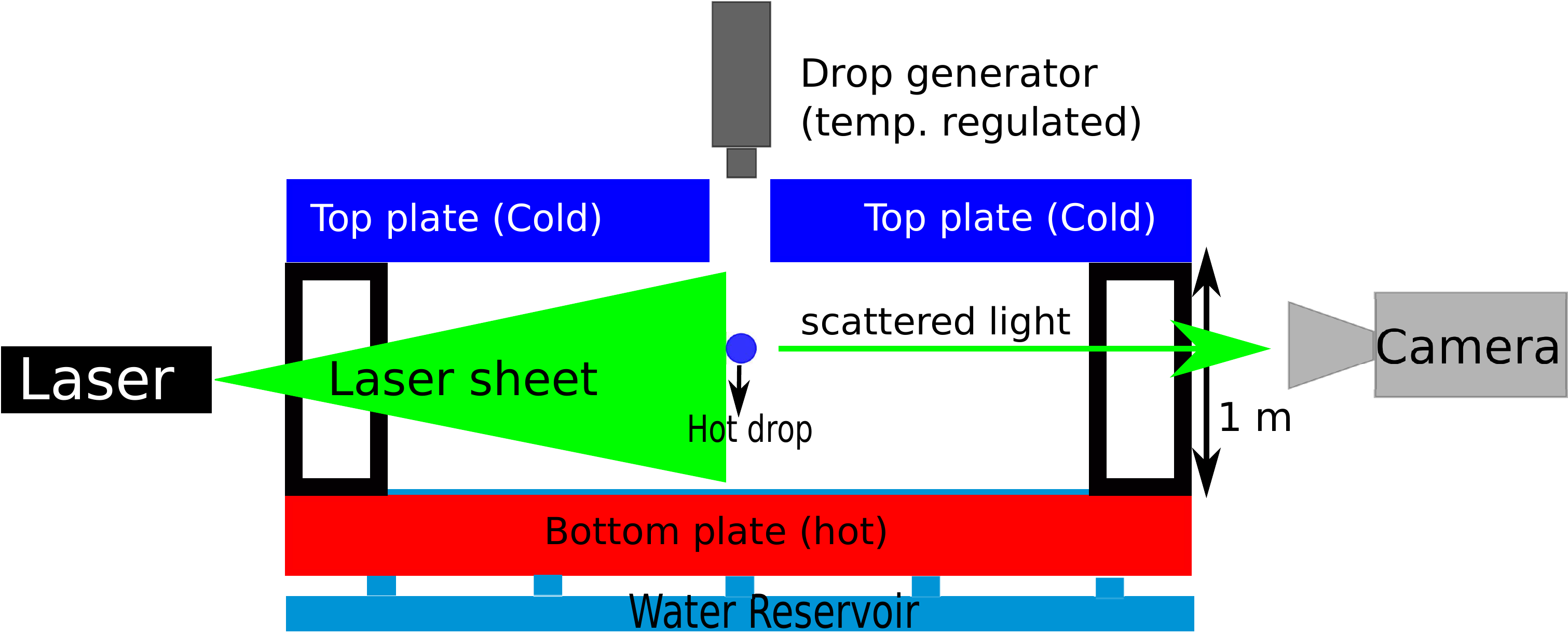}
\caption{Layout of the experimental setup}
\label{fig:exp_layout}
\end{figure}

%Materials and Methods\\
\textit{Experimental Setup}\\
The experiments were conducted in the turbulent cloud chamber facility at Michigan Tech. \cite{chang2016laboratory}. The chamber is a standard Rayleigh-B\'enard convection experiment driven by an unstable thermal gradient imposed between the top and bottom plates. The layout of the experimental setup is shown in Fig.~\ref{fig:exp_layout}. The top and bottom plates were covered with moist filter paper. The side wall conditions were adjusted such that the relative humidity in the chamber was less than $100\%$. Under cold conditions, the top and bottom plates were set to the same temperature to avoid the loss of water vapor from the bottom plate. The drop generator produced 2 mm diameter drops at a set temperature every few seconds. The wake of the drop was illuminated using a vertical, planar light sheet from an MGL-III $532$\,nm $300$\,mW continuous laser. The forward scattered light from the nucleated particles was recorded using a Sony alpha 7S II camera placed across the chamber. The experiments were conducted under warm and cold conditions using Sodium Chloride and aqueous Snowmax as aerosols respectively, produced using a TSI atomizer (Model 3076).

\begin{figure}[h]
\centering
\includegraphics[height = 0.9\textheight]{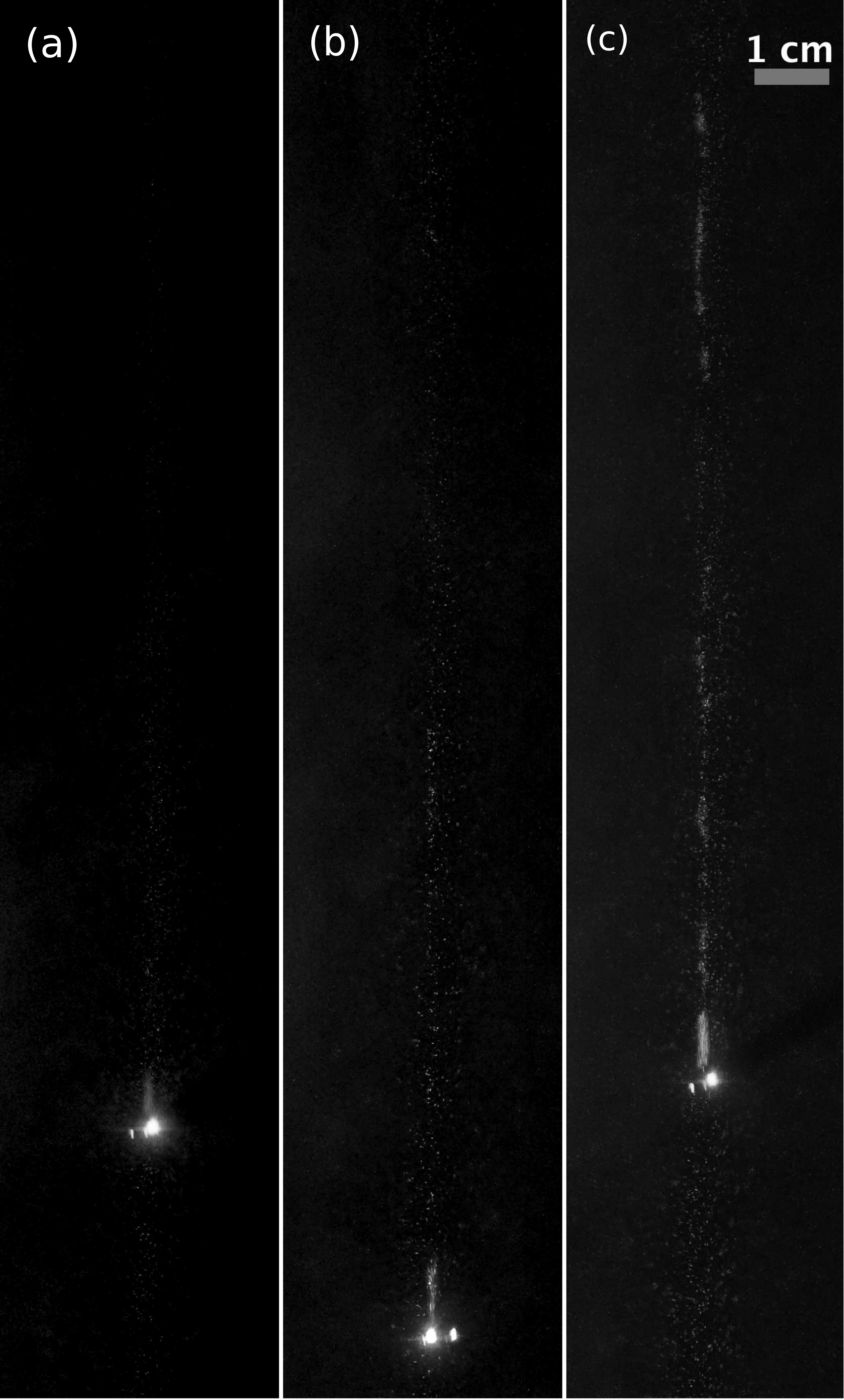}
\caption{Nucleation of water droplets in the wake of a hot drop under warm conditions. Ambient temperature $\approx~11~^{\circ}$C, humidity $\approx~92\%$. The initial drop temperatures are (a) $25~^{\circ}$C (b) $30~^{\circ}$C (c) $35~^{\circ}$C. The number concentration of the nuclei is about $1.8\times10^4$ per cm$^{3}$. The images are contrast adjusted.}
\label{fig:warm_cloud}
\end{figure}

\begin{figure}[h]
\centering
\includegraphics[width = \textwidth]{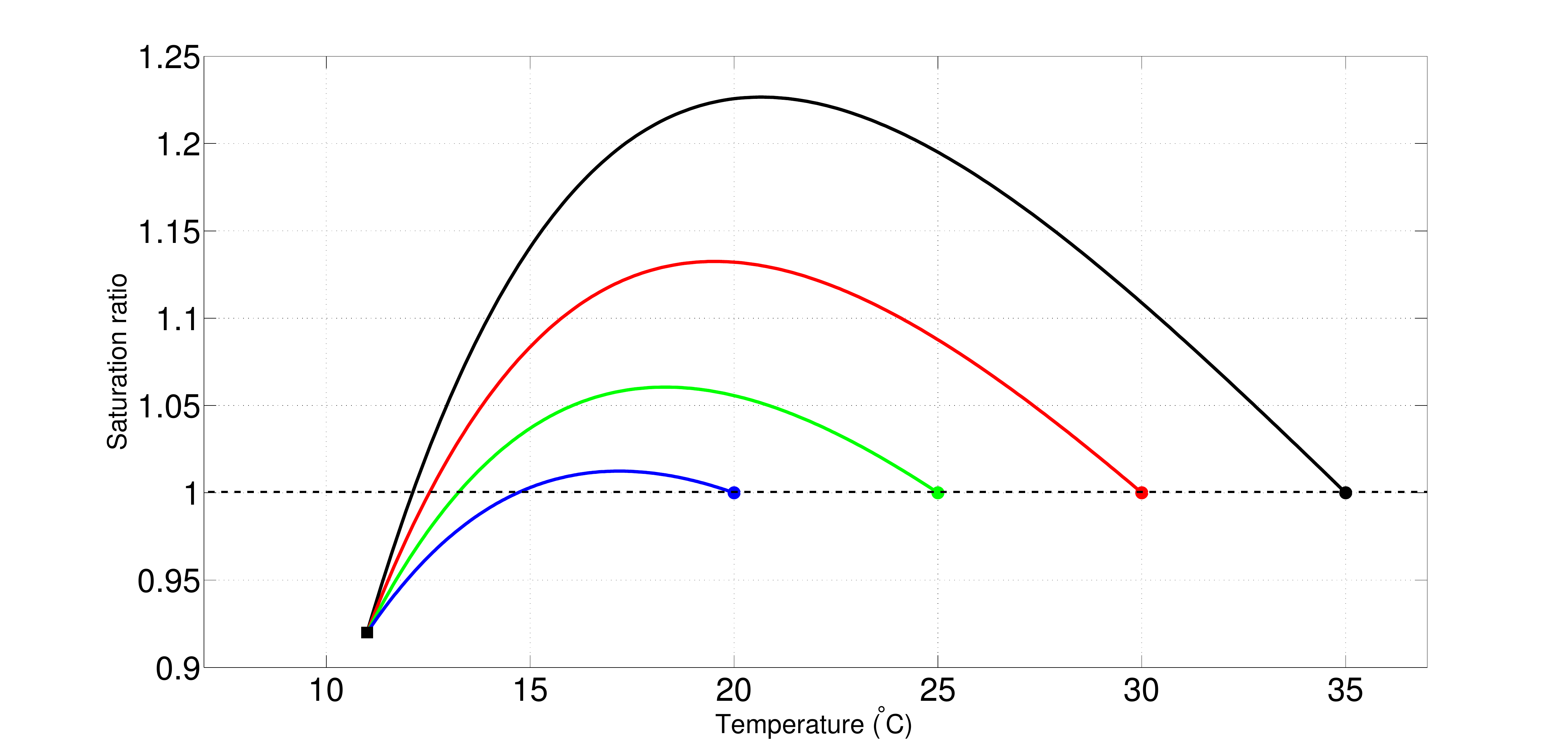}
\caption{Saturation ratio using the mixing parcel model for the experiments in Fig.~\ref{fig:warm_cloud}. The conditions at the surface of the drop is marked in filled circles ($\bullet$) of various colors and the corresponding mixing line is represented by a curve connecting $\bullet$ (of different colors) and $\blacksquare$: $\bullet$ - $35~^\circ$C, $\textcolor{red}{\bullet}$ - $30~^\circ$C, $\textcolor{green}{\bullet}$ - $25~^\circ$C,$\textcolor{blue}{\bullet}$ - $20~^\circ$C. The black horizontal line indicates saturation ratio = 1.}
\label{fig:warm_SS}
\end{figure}

Supplementary Text\\
Figure~\ref{fig:warm_cloud} shows the dynamics observed in the wake of a falling hot drop at various drop temperatures. The temperature of the drop increases as we move from left to right in Fig.~\ref{fig:warm_cloud}. The bottom and top plates along with the side walls were set to the desired temperatures such that the average humidity inside the chamber was below water saturation. From Fig.~\ref{fig:warm_cloud}, we infer that the spatial extent of the nucleated droplets in the wake is a strong function of the drop temperature. Similar to the experiments under cold conditions (see main text), we observe that the spatial extent of the nucleated droplets in the wake increases as the temperature difference between the drop and the ambient was increased. Figure~\ref{fig:warm_SS} shows the corresponding variation of the saturation ratio obtained for different drop temperatures. Saturation ratio is defined as  $S=P_w/P_s(T_w)$ where $P_s(T)$ is the saturated vapor pressure at a temperature $T$ and $P_w$ is the vapor pressure in the wake obtained from the parcel mixing model (see main text for details on the mixing model).

\end{document}